# Three possible mechanisms of capacitance enhancement under magnetic field: charge density gradient modulation, electron gas excitation and oscillatory magnetization-polarization coupling

Ye Wu, Ruyan Guo and Amar Bhalla.

The University of Texas at San Antonio, Department of Electrical and Computer Engineering, One UTSA Circle, San Antonio, TX 78249.

E-mail: ruyan.guo@ utsa.edu

**Abstract**

Three mechanisms of capacitance enhancement by a magnetic field have been analyzed. Through semiclassical description of charge movement under a magnetic field, it can be shown that the charge density gradient, a term coupled with magnetic field strength, could be used to augment the magnetocapacitance. Also a magnetic field could enhance capacitance on the polarized metallic material surface due to the electron gas excitation. Finally, a magnetic field could produce oscillation in the capacitance when relating the polarization with the magnetization through the modification of standard free energy model. By deriving these three mechanisms, it can be seen that three approaches are of potential for exploring tunable dielectric materials.

## 1.Introduction

Magnetically tunable dielectric material is typically thought to be produced by magneto-electric coupling. It has been shown that the magnetic moment could be generated by the spin current in the so called spiral spin structure and therefore manipulating the spontaneous polarization[1]. One manifestation of such manipulation could be the polarization introduced in $SmFeO_3$ single crystal [2]and $DyMnO_3$ thin film[3]. The correlation of the polarization response with magnetization has established the assumption that we may tune magnetization plus the polarization through the magnetic field and thus may expect viable behavior of the capacitance. Another possible approach is the doping of the functional guest magnetic molecule into the host material. For example, the researchers claim the capacitance enhancement of the whole nanocomposites through using $Fe_2O_3$ to decorate liquid crystal and grapheme [4] [5].

Although these two approaches are distinct in material design, the overall mechanisms are almost the same. It involves the repulsive and attractive Coulomb interaction in a magnetic field and the polarization–magnetization coupling. Therefore these facts establish that we may be able to tune the capacitance through the external magnetic field once we find the tunable parameters involving with charge density or magnetization.

Based on this approach, we examine the three mechanisms of capacitance enhancement under magnetic field. Firstly, we investigate the correlation between the charge density profile and the variation of the magnetic field. From this, we deduce that the capacitance is depended on the magnetic field strength and the charge density. Secondly, by assuming that the electron gas excitation could lead to the charge rearrangement on the metallic material surface, we calculate the capacitance of the surface layer with respective to the magnetic field strength. Finally, we find that the magnetization and the electric polarization response of dielectric material both show magnetically oscillatory behavior, which leads to the oscillatory enhancement of capacitance under the magnetic field. A complete confirmation of the mechanisms would be the future involvement in experimental investigations of the magnetic field enhancement of the capacitance, which would lead to the full understanding of the magnetocapacitance modulation.

## 2. Effect of the charge density gradient on the capacitance.

From the basic connection between capacitance and the charge distribution[6], one gets

$$C = eA\frac{\partial n}{\partial V} ,  \quad (2.1)$$

where $e$ is unit electron charge, $A$ is the interfacial area, $n$ is the electron charge density, $V$ is the localized electric potential. We know that the electron would undergo a magnetic force and would travel with a specific distance. Based on this fact, we are able to explore the consequences of the change in charge density induced by an applied magnetic field. We would use a semiclassical method [7] to describe the charge movement in a magnetic field. As shown in Appendix A, The capacitance could be derived as

$$C = \frac{6C_0 H_x \sqrt{(C_1 n^{5/3} - C_2)(C_3 n^{1/3} - C_4)}}{9n^{-5/3}\sqrt{C_1 n^{5/3} - C_2} - 4n^{-1/3} H_x \sqrt{C_3 n^{1/3} - C_4}} .$$

(2.2)

where $C_1, C_2, C_3$ and $C_4$ are the constants defined in Eq.(A.18), Eq.(A.19), Eq.(A.20) and Eq.(A.21), $n$ is the charge density and $H_x$ is the magnetic field strength along x-axis.

Figure 1 shows the capacitance dependence on the magnetic field and charge density, which could be estimated from Eq.(2.2). These results suggest quasi-linear enhancement of capacitance by magnetic field (Figure 1(a)) and the larger capacitance is associated with the increasing of the charge density and magnetic field. This finding illustrates the importance of using doping to alter the charge distribution in solid material which could lead to the capacitance enhancement.

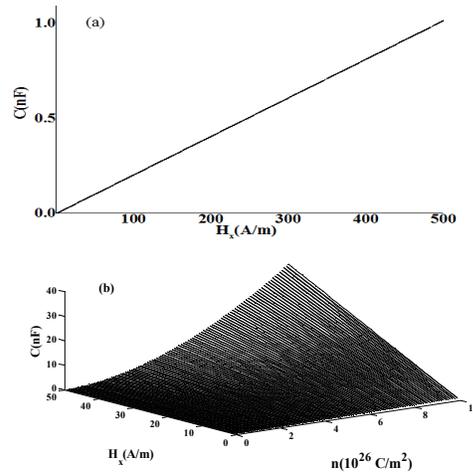

**Figure 1.** (a) Simulation of capacitance with respect to magnetic field strength. (b) The capacitance with respect to the magnetic field strength and the charge density.

## 3. Capacitance enhancement due to the charge rearrangement on the metallic material surface.

The nature of the capacitance modulation is a subject of ongoing discussions. The above section and Appendix A demonstrates that the magnetic force could introduce charge density gradient in the solid, which is one way of charge rearrangement in essence. This result suggests that the charge rearrangement may be potentially used for the capacitance modulation. For the following, let us consider one typical phenomenon of the charge rearrangement, which is the electron gas excitation that happens on the metallic material surface. From the fact that the collective excitations, such as plasmons, could cause the charge rearrangement, one may expect that under magnetic field radiations, the capacitance on the metallic surface could be modulated to a certain extent. As shown in Appendix B(Eq.(B.17)), we derived the expression of capacitance,

$$C=\frac{2C_0C_5e^2At^{2/5}H_x^{3/5}}{3m\varepsilon_0(A_1A_2)^{3/5}}.\quad (3.1)$$

which shows that the capacitance on the metallic material surface is dependent on the magnetic field strength $H_x$. As shown in Figure 3 (a), it predicts that the capacitance is enhanced by the magnetic field on the metallic material surface.

For the following, it will be important to know how the electron gas excitation would affect the capacitance on the metallic material surface. We would start from the plasma frequency, a key parameter associated with the electron gas excitation. Using Eq.(B.6) and Eq.(B.11) in Appendix B, the capacitance could be rewritten as

$$C=\frac{3C_0C_5Am\varepsilon_0\omega^2}{4C_6e^2\cos(\omega t-\varphi)},\quad (3.2)$$

It illustrates that the capacitance could be tuned by the plasma frequency. Figure 2(b) plots the oscillatory modulation of the capacitance by the plasma frequency. This confirms our assumption that the electron gas excitation would affect the capacitance on the metallic material surface. This indicates the second possible mechanism that the electron gas excitation, a way of charge rearrangement on the metallic surface, could lead to capacitance enhancement under a magnetic field.

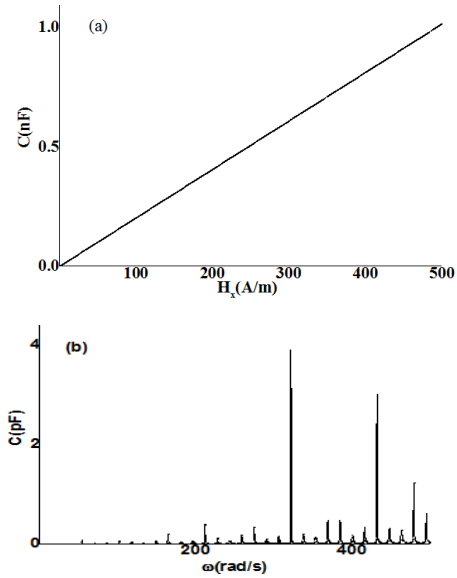

Figure 2. (a) The capacitance enhanced by the magnetic field strength on the metallic material surface.(b)The capacitance tuned by the plasma frequency.

## 4. de Haas–van Alphen Oscillations in a dielectric material at low magnetic field and room temperature.

Generally, the magnetization could show oscillatory behavior when the magnetic field is applied, which is known as the de Haas-van Alphen (dHvA) effect[8]. This effect is mostly shown in organic superconductor under strong magnetic field and at low temperature. [9]

In some dielectric material, the coupling between the electric polarization and magnetization is very strong, which is generally described as the magneto-electric effect. The correlation of polarization response with magnetization has established the assumption that we may tune magnetization plus the polarization through the magnetic field and may expect oscillatory enhancement of the magnetocapacitance in dielectric material.

Let us suppose that the free energy of the whole system involves with interaction between the polarization and the electrical field, the coupling between the magnetization and the polarization. We modify the LEI model [10] and consider that the whole free energy is

$$F = F_0 + 2\pi/P^2\varepsilon_0 + C_0 P^4 - E_{efi} P - E_{efi}^2/8\pi + P^2 M^2$$ (4.1)

where $F_0$ and $C_0$ are constants, $E_{efi}$ is the electrical field intensity, $M$ is the magnetization and $P$ is the electrical polarization. In this equation, the first five terms are cited directly from LEL model[10], which state the interaction between the polarization and the electrical field. We add the last term $P^2 M^2$ that could represent the coupling between the magnetization and the polarization. Its precise format could be $P^2 \langle M_q / M_{-q} \rangle$, which is dependent on the wave vector $q$ [11]. Here we only used its approximation

$$P^2 \langle M_q / M_{-q} \rangle \approx P^2 M^2 . \text{ [11]}$$ (4.2)

The energy could have its minimum when the system is in the equilibrium state. By taking the extreme, we have

$$\partial F / \partial P = 4C_0 P^3 + (4\pi/\varepsilon_0 + 2M^2)P - E = 0.$$ (4.3)

If we let the constant $C_0 = 1$, we have the real root,

$$P_1 = \sqrt[3]{-E_{efi}/8 + \sqrt{E_{efi}^2/64 + (4\pi/\varepsilon_0 + 2M^2)^3/1728}}$$
$$+ \sqrt[3]{E_{efi}/8 - \sqrt{E_{efi}^2/64 - (4\pi/\varepsilon_0 + 2M^2)^3/1728}}$$ (4.4)

If we let $C_0 = 0$, we have another solution,

$$P_2 = E_{efi}/(2M^2 + 4\pi/\varepsilon_0),$$ (4.5)

It is noted that Eq.(4.5) is similar to the result of Eq.(3) in Ref.[11].

We suppose the ultimate solution to the polarization is the superposition of the two polarization $P = C_1 P_1 + C_2 P_2$, (4.6)

where $C_1, C_2$ are constants.

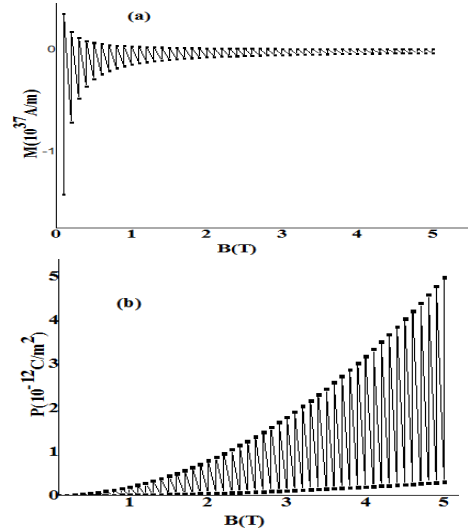

**Figure 3**. (a) Magnetization vs. the magnetic field. (b) Polarization vs. the magnetic field.

We have derived the polarization dependence on the magnetization in Eq.(4.4) and Eq.(4.5). In order to show the oscillating behavior of capacitance, the proceeding would be the derivation of the oscillating magnetization. The detailed derivation is shown in Appendix C. The expression of the magnetization is shown in Eq.(C.11),Eq.(C.12) and Eq.(C.13).

Interestingly, the magnetization could switch to negative (Figure 3(a)), which suggests the magnetic moment reverses and becomes antiparrallel to external fields. Eq.(C.12) and Eq.(C.13) are the oscillating components of the magnetization. Plugging all the magnetization components into Eq.(4.4) and Eq.(4.5), we can calculate the oscillating polarization(Figure 3(b)) associated with the magnetization. The capacitance could be further simulated using the value of polarization. It is noted that the capacitance modulation could be affected by harmonic oscillation numbers. As shown in Figure 4, with the more harmonic oscillation involved, the capacitance is much more enhanced by the magnetic field.

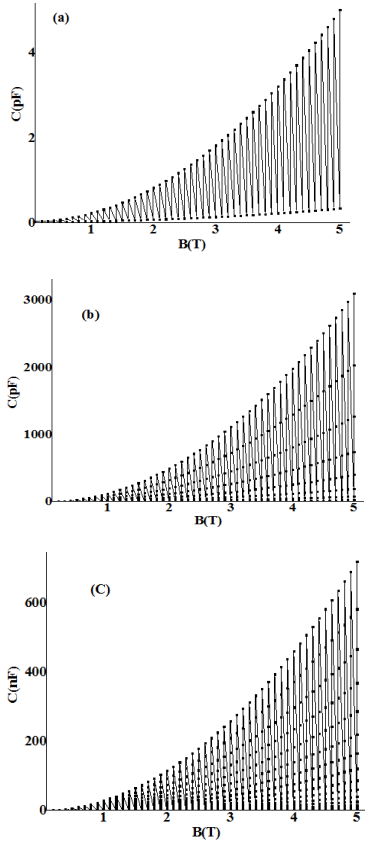

**Figure 4**. Oscillatory magnetocapacitance with varied harmonic oscillation number (a) $1 \leq m \leq 2$ (b) $1 \leq m \leq 10$ (c) $1 \leq m \leq 40$

## 5. Conclusion.

We have theoretically shown the effect of a magnetic field on the capacitance. Three mechanisms are used to explore this magnetocapacitance tunability. In the first mechanism, the charge density gradient is used to describe the magnetocapacitance through the basic electrodynamics. The relation between the charge density gradient and the magnetic field strength is derived, which leads to final expression of capacitance dependence on the magnetic field. This mechanism predicts a quasi-linear dependence of the capacitance on the magnetic field strength. In the second mechanism, the electron gas excitation is found to enhance the magnetocapacitance on the metallic surface. In the third mechanism, the dHvA effect involved with the dielectric response is examined. On expanding the oscillatory magnetization and coupling it with the polarization through the classical expression of the free energy, we simulated the oscillatory polarization under magnetic field, which leads to oscillatory and tunable capacitance. Here, the simulated result of oscillatory magnetocapacitance suggests a very distinct dHvA phenomenon under low magnetic field (less than 2T) and at room temperature, which needs further experimental confirmation. The three possible mechanisms derived here show three possible approaches to tunable capacitance and suggest a perspective for the design of tunable dielectric material and devices.

### Acknowledgments

This work was supported by the NSF and INAMM. The authors thank Prof.Robert Whetten and Prof.Changhong Yi at UTSA Physics Dept. for their many fruitful discussions on the topics presented here.

**Appendix A:**

It should be noted that the derivation of the charge density gradient in the following is similar to Ref.[15]. The key step is using the classical force balance equation to derive the relation between the charge density and magnetic field. For length consideration, we only show some key equations and concepts.

Assuming the magnetic field is along the x-direction and ignoring the velocity in the z-direction due to the equilibrium state, the magnetic force would be derived as

$$F_y = eH_x v_z. \quad (A.1)$$

For the following, we will derive the attractive force and repulsive force. Using the Poisson's equation,

We can derive the attractive force as,

$$e(E_0 + \tau e \int_0^y n(\gamma) d\gamma). \quad (A.2)$$

where $\tau$ is the parameter that depends on the material permittivity and could be fitted by the experimental data, e is the elementary charge, $n$ is the charge density

The repulsive force could be expressed as

$$\frac{\hbar^2}{m} \pi^{4/3} 3^{-1/3} n^{-1/3} \frac{dn}{dy}. \quad (A.3)$$

Then the force balance equation is given by

$$\frac{\hbar^2}{m} \pi^{4/3} 3^{-1/3} n^{-1/3} \frac{dn}{dy} = e(E_0 + ke \int_0^y n(\gamma) d\gamma) + eH_x v_z. \quad (A.4)$$

Considering that the induced magnetic field

$$H_i = 4\pi \chi(y) H \quad (A.5)$$

the total magnetic field is given by

$$H_t = H_i + H \quad (A.6)$$

Using Landau diamagnetism, the susceptibility is given by

$$\chi(y) = \frac{-e^2 k_F}{12\pi^2 mc^2} = \frac{-e^2}{12mc^2} 3^{1/3} \pi^{-4/3} n(y)^{1/3}. \quad (A.7)$$

Using Maxwell equation

$$\nabla \times \vec{H}_t = \partial \vec{D}/\partial t + \vec{J}, \quad (A.8)$$

we have $-\partial H_t / \partial y = J_z$, (A.9)

Since the local current is given by

$$J_z = e \cdot n \cdot v_z, \quad (A.10)$$

where $v_z$ is the local velocity of the charge. then the velocity is given by

$$v_z(y) = \frac{-e 3^{1/3} \pi^{-4/3} H_x n(y)^{-5/3}}{12 mc^2} \frac{\partial n(y)}{\partial y}. \quad (A.11)$$

Reorganizing the force balance equation again, then

$$eE_0 + ke^2 \int_0^y n(\gamma) d\gamma = \frac{e^2 3^{1/3} \pi^{-4/3} H_x^2 n(y)^{-5/3}}{12 mc^2} \frac{dn(y)}{dy} + \frac{\hbar^2}{m} \pi^{4/3} 3^{-1/3} n(y)^{-1/3} \frac{dn(y)}{dy}, \quad (A.12)$$

Using the boundary condition $\frac{dn}{dy}\big|_{y=a}=0$

(A.13)

and $n\big|_{y=a}=n_a$, (A.14)

The charge density could be shown to be,

$$y(n) = \int_n^{n_a} \frac{2\xi^{-1/3}d\xi}{3\sqrt{C_1\xi^{5/3}-C_2}}$$
$$-\int_n^{n_a} \frac{3\xi^{-5/3}d\xi}{2\sqrt{C_3\xi^{1/3}-C_4}}\frac{1}{H_x},$$

(A.15)

where

$$C_1 = \frac{8\pi e^2 m}{15\hbar^2\pi^{4/3}3^{-1/3}},$$ (A.16)

$$C_2 = \frac{8\pi e^2 m a^{1/3}}{\hbar^2\pi^{4/3}3^{2/3}},$$ (A.17)

$$C_3 = \frac{2\pi e^4 3^{-5/3}\pi^{-4/3}}{mc^2},$$ (A.18)

$$C_4 = \frac{2\pi e^4 3^{-5/3}\pi^{-4/3}a^{1/3}}{mc^2}.$$ (A.19)

The capacitance in Eq.(2.1) could be further derived as

$$C = eA\frac{\partial n}{\partial V} = eA\frac{\partial n}{E_L(\omega)\partial y} = C_0\frac{\partial n}{\partial y},$$

(A.20)

where $E_L(\omega)$ is the localized electric field that is dependent on the electric signal frequency, $C_0$ is the parameter which is depended on $E_L(\omega)$ and could be fitted from the experimental data. This shows that the capacitance is subjected to charge density gradient modulation.

Taking the integral in both sides of Eq.(A.15), the charge density gradient is given by

$$\frac{\partial n}{\partial y} = \frac{-6H_x\sqrt{C_1n^{5/3}-C_2}\sqrt{C_3n^{1/3}-C_4}}{-4H_x n^{-1/3}\sqrt{C_3n^{1/3}-C_4}+9n^{-5/3}\sqrt{C_1n^{5/3}-C_2}}.$$

(A.21)

Using the value of $\partial n/\partial y$, $C$ is derived as

$$C = \frac{6C_0 H_x\sqrt{(C_1n^{5/3}-C_2)(C_3n^{1/3}-C_4)}}{9n^{-5/3}\sqrt{C_1n^{5/3}-C_2}-4n^{-1/3}H_x\sqrt{C_3n^{1/3}-C_4}}.$$

(A.22)

**Appendix B**:

Let us consider a metallic bulk subjected to a magnetic field is polarized to generate a localized electric field. It is assumed that all the polarization occurs at the surface. The electrons are supposed to protrude a certain distance $l$ on one side and the positive nuclei protrude a distance $l$ on the other side. It is noted that the local charge is given by
$Q_L = nel$, (B.1)

where $n$ is the charge density and $e$ is the unit charge. Using Gauss's Law, the localized electric field is given by
$E_L = Q_L/\varepsilon_0 = nel/\varepsilon_0$. (B.2)

Applying the Newton's Law, the force introduced by the electric field is expressed as $F_L = -eE_L = -ne^2l/\varepsilon_0 = ma = m\ddot{l}$. (B.3)

It is supposed that the plasma oscillations appear on the surface of this metallic bulk. Furthermore, the equation of the electron gas harmonic motion is expressed as

$$-kl = m\ddot{l} \quad (B.4)$$

and the solution is $l = A\cos(\omega t - \varphi)$. (B.5)

Therefore the plasma oscillation frequency could be derived as $\omega = \sqrt{\dfrac{ne^2}{m\varepsilon_0}}$, (B.6) and the velocity as

$$\dot{l} = -\omega A \sin(\omega t - \varphi). \quad (B.7)$$

For the following, we would correlate the distance $l$ with the magnetic field. As shown in the Section 2, the velocity that due to the magnetic force could be expressed as

$$v = \frac{-e3^{1/3}\pi^{-4/3}H_x n(l)^{-5/3}}{12mc^2}\frac{\partial n(l)}{\partial l}. \quad (B.8)$$

Taking the integral in both sides of Eq.(B.8) and reorganizing it, we obtain

$$\frac{C_5 v l}{H_x} = n^{-2/3}, \quad (B.9)$$

where $C_5 = \dfrac{8mc^2}{e3^{1/3}\pi^{-4/3}}$. (B.10)

It is noted that $\dfrac{C_6 \omega \sin(2\omega t - 2\varphi)}{H_x} = n^{-2/3}$,

(B.11)

where $C_6 = -C_5 A^2 / 2$. (B.12)

Plugging Eq.(B.6) into Eq.(B.11), we further obtain $A_1 n^{7/6}\sin(A_2 \sqrt{n} t - 2\varphi) = H_x$,

(B.13)

It is assumed that the angle of $(A_2 \sqrt{n} t - 2\varphi)$ is small and the initial phase angle $\varphi = 0$. (B.14)

Then using Taylor series expansion, we can derive $n \approx \left(\dfrac{H_x}{A_1 A_2 t}\right)^{3/5}$. (B.15)

Similarly we can derive

$$v = -\omega A \sin(\omega t - \varphi) \approx -\omega^2 A t = -\frac{ne^2 A t}{m\varepsilon_0}.$$

(B.16)

Using Eq.(B.15) and (B.16), the capacitance could be expressed as

$$C = C_0 \frac{1}{\partial l / \partial n} = \frac{C_0 C_5 v}{(-2/3)n^{-5/3}H_x} \approx \frac{2C_0 C_5 e^2 A t^{2/5} H_x^{3/5}}{3m\varepsilon_0 (A_1 A_2)^{3/5}}$$

(B.17)

**Appendix C**:

The physical mechanism which is responsible for the dHvA effect is the energy splitting in the magnetic field. In order to confirm our argument that due to the dHvA effect, the capacitance could show magnetically oscillatory response, we have to show the capacitance or the polarization association with energy level. We have to consider the energy involvement in the magnetization as the following. Similar to the classical treatment of oscillatory magneto-resistance dependence on the magnetic field, we begin from the derivation of the oscillating components of the magnetization. Firstly, we have to derive the oscillatory part of the thermodynamic potential.

Initially, the potential of the whole system is given by [8]

$$\Omega = -\sum_{n=0}^{\infty} T \ln[1 + e^{(\mu - E_n)/T}], \quad (C.1)$$

where $\mu$ is the chemical potential. $T$ is the environmental temperature. $E_n$ is the electron energy level in the magnetic field.

The Laudau energy level in magnetic field is solved as [12]

$$E_n = \frac{\hbar e B}{m^* c}(n + \frac{1}{2}) + \frac{\hbar^2 k_0^2}{2m^*}, \quad (n = 0,1,2,3,\ldots) \quad (C.2)$$

where $\hbar$ is the Planck's constant, $e$ is the electron charge, $B$ is the external magnetic field, $k_0$ is the wave vector magnitude. $m^*$ is the electron mass.

The chemical potential could be solved using Hubbard model [13], which lead to the solution of

$$\mu = -2t \cos(\frac{N\pi}{2})\sqrt{1 - \frac{\mu_B^2 B^2}{4t^2 \sin^2(N\pi/2)}}, \quad (C.3)$$

where $t \approx -\pi E_n / 2 / \sin(\pi N) \quad (C.4)$

is the electron hopping rate in the fully polarized case with all the electron spins aligned, $N$ is the electron density, $\mu_B$ is the Bohr magneton, $B$ is the external magnetic field.

We firstly substitute Eq.(4.8) and (4.9) into Eq. (4.7), and then expand the potential with respective to $B$ through using Taylor series expansion and the Poisson summation formula, [14]

$$f(n) = f(0)/2 + \int_0^{\infty} dn f(n) + 2\text{Re} \int_0^{\infty} dn f(n)$$
$$+ 2\text{Re} \sum_{k=1}^{\infty} \int_0^{\infty} dn f(n) \exp(i2\pi k n) \quad (C.5)$$

Then we can derive the potential as
$$\Omega = \Omega_1 + \Omega_2 + \Omega_3, \quad (C.6)$$

Where

$$\Omega_1 = -T \ln\{1 + \exp[-\frac{2t}{T}\cos(\frac{N\pi}{2})$$
$$\sqrt{1 - \left(\frac{\mu_B^2 B^2}{4t^2}\right)\frac{1}{\sin^2(N\pi/2)}} - \frac{\hbar e B}{2m^* c} - \frac{\hbar^2 k^2}{2m^*}]\},$$

(C.7)

$$\Omega_2 = \sum_{m=1}^{\infty}(-1)^m (\frac{T^2 m^* c}{\hbar e B m^2}) \exp[\frac{e B \hbar m}{m^* cT} - \frac{2mt}{T}\cos(\frac{N\pi}{2})$$
$$\sqrt{1 - \frac{\mu_B^2 B^2}{4t^2 \sin^2(N\pi/2)}} - \frac{\hbar^2 k_0^2 m}{2m^* T}],$$

(C.8)

$$\Omega_3 = \sum_{k=1}^{\infty} \sum_{m=1}^{\infty} \frac{(-1)^{m+1} 2m^* c T m B^2}{(\hbar e m B)^2 + (2\pi k m^* cT)^2}$$
$$\exp[-(\frac{2mt}{T})\cos(\frac{N\pi}{2})\sqrt{1 - \frac{\mu_B^2 B^2}{4t^2 \sin^2(N\pi/2)}}.$$
$$-\frac{\hbar^2 k_0^2 m}{2m^* T}]$$

(C.9)

By taking the derivative over $B$, we obtain the magnetization as

$$M = -\partial \Omega / \partial B = M_1 + M_2 + M_3, \quad (C.10)$$

where

$$M_1 = -(\frac{T}{2})\exp[-2t\cos(\frac{N\pi}{2})\sqrt{1-\frac{\mu_B^2 B^2}{4t^2\sin^2(N\pi/2)}} - \frac{c\hbar eB}{2m^*} - \frac{\hbar^2 k_0^2}{2m^*}]/\{1+\exp[-\frac{2t}{T}\cos(\frac{N\pi}{2})\sqrt{1-\frac{\mu_B^2 B^2}{4t^2\sin^2(N\pi/2)}} - \frac{\hbar eB}{2m^*c} - \frac{\hbar^2 k_0^2}{2m}]\}$$

(C.11)

$$M_2 = \sum_{m=1}^{\infty}\frac{(-1)^m T^2 m^* c}{\hbar em^2 B^2}\exp[\frac{eB\hbar m}{m^* cT} - \frac{2mt}{T}\cos(\frac{N\pi}{2})\sqrt{1-\frac{\mu_B^2 B^2}{4t^2\sin^2(N\pi/2)}} - \frac{\hbar^2 k_0^2 m}{2m^* T}] + \sum_{m=1}^{\infty}\frac{(-1)^m T^2 m^* c}{\hbar emB}[\frac{e\hbar}{m^* cT} + \frac{\mu_B^2 B}{T\tan(N\pi/2)}\frac{1}{\sqrt{4t^2\sin^2(N\pi/2)-\mu_B^2 B^2}}]\{\exp[\frac{eB\hbar m}{m^* cT} - \frac{2mt}{T}\cos(\frac{N\pi}{2})\sqrt{1-\frac{\mu_B^2 B^2}{4t^2\sin^2(N\pi/2)}} - \frac{\hbar^2 k_0^2 m}{2m^* T}]\}$$

(C.12)

$$M_3 = \sum_{k=1}^{\infty}\sum_{m=1}^{\infty}\frac{(-1)^m 16\pi^2 k^2 m^{*3} c^3 T^3 \hbar emB}{(\hbar^2 e^2 m^2 B^2 + 4\pi^2 k^2 m^{*2} c^2 T^2)^2}\{\exp[\frac{-2mt\cos(N\pi/2)}{T}\sqrt{1-\frac{\mu_B^2 B^2}{4t^2\sin^2(N\pi/2)}} - \frac{\hbar^2 k_0^2 m}{2m^* T}]\} + \sum_{k=1}^{\infty}\sum_{m=1}^{\infty}\frac{(-1)^m 2\hbar em^* cm^2 \mu_B^2 B^3}{(\tan(N\pi/2))(\hbar^2 e^2 m^2 B^2 + 4\pi^2 k^2 m^{*2} c^2 T^2)}\frac{1}{\sqrt{4t^2\sin^2(N\pi/2)-\mu_B^2 B^2}}\{\exp[\frac{-2mt}{T}\cos(N\pi/2)\sqrt{1-\frac{\mu_B^2 B^2}{4t^2\sin^2(N\pi/2)}} - \frac{\hbar^2 k_0^2 m}{2m^* T}]\}$$

(C.13)